\documentclass[aps,prb,twocolumn,superscriptaddress,showpacs,superbib,floats]{revtex4}
\usepackage{graphicx}
\usepackage{amssymb}
\usepackage{amsmath}
\usepackage{dcolumn}
\usepackage{color}

\begin{document}

\title{Vibrational and Thermal Properties of ZnX (X=Se, Te): Density Functional Theory (LDA and GGA) versus Experiment}

\author{R. K. Kremer}
\email[Corresponding author:~E-mail~]{R.Kremer@fkf.mpg.de}
\author{M. Cardona}
\author{R. Lauck}
\author{G. Siegle}
\affiliation{Max-Planck-Institut f{\"u}r Festk{\"o}rperforschung,
Heisenbergstrasse 1, D-70569 Stuttgart, Germany}

\author{A.H. Romero}
\affiliation{CINVESTAV, Departamento de Materiales, Unidad
Quer$\acute{e}$taro, Quer$\acute{e}$taro, 76230, Mexico}

\date{\today}

\begin{abstract}
We calculated the phonon dispersion relations of ZnX (X=Se, Te) employing \textit{ab initio} techniques. These relations have been used to evaluate the temperature dependence of the respective specific heats of crystals with varied isotopic compositions. These results have been compared with measurements performed on crystals down to 2 K. The calculated and measured data are generally in excellent agreement with each other. Trends in the phonon dispersion relations and the corresponding densities of states for the zinc chalcogenide series of zincblende-type materials are discussed.

\end{abstract}

 \maketitle


\email{R.Kremer@fkf.mpg.de}

\section{Introduction}

In an ongoing effort to characterize the vibrational and thermal properties of multinary semiconductors and semimetals of current interest we have recently focused our attention on binary and ternary chalcogenides, such as HgX (X=S, Se, Te), CuGaS$_2$ or ZnS in the zinc-blende (zb)  and rocksalt structure.\cite{Cardona2009,Romero2008,Cardona2010b,Romero2011,Cardona2010a} Most of our past experimental work has been concerned with the specific heats in the low-temperature regime where deviations from the Debye-$T^3$ power law can be easily studied.
Such deviations arise from low-energy phonon branches with low dispersion, typically zone-boundary transverse acoustic (TA) phonon branches.
\textit{Ab initio} calculations of the specific heat ($C_v$) based on the electronic structure derived from local density functionals have been carried out and compared in detail with the experimental results.  They were found to account rather well for the experimental data and, in some cases, confirm our experimental results where large deviations from previously published measurements were encountered,  e. g. for GaN.\cite{Kremer2005}
For most of our investigated systems, by selective isotope substitution we could effect incremental variations of the low-temperature heat capacities.\cite{Gibin2005} The small changes of the heat capacities induced by selective isotope substitution are generally very well described by density functional theory (DFT) and were advantageously used to contrast the calculations with experimental data. The \textit{ab initio} calculations were done employing the two most common approximations for the density functional of the exchange and correlation energy, the local density approximation (LDA) and the generalized gradient approximation (GGA) implemented in various DFT codes. Usually, LDA and GGA results bracket the experimental data. For example, LDA tends to underestimate the lattice parameters and the cell volume, while the GGA approximation overestimates them. This has lately been discussed by us, for example, for ZnS, the cubic mercury chalcogenides HgX (X=S, Se, Te) and  cinnabar ($\alpha$-HgS). For systems including heavy elements we have also tested the effect of including spin-orbit (SO) interaction into the DFT calculations. For $\alpha$-HgS, SO coupling was found to give a negligible effect on the thermal properties. However, the heat capacity and also the phonon dispersion relations of the semimetal Bi can be noticeably better described if SO interaction is included in the DFT calculations.\cite{Diaz2007} SO interaction effects on the thermal properties are less pronounced for Sb, another semimetal.\cite{Serrano2008} The relativistic electronic structures which
results from our calculations enabled us to study a number of interesting band structure effects such as linear SO splittings of the valence band  near the origin of the Brillouin zone.\cite{Cardona2009,Cardona2010b} In addition to this program, also various DFT implementations (ABINIT and VASP) have been compared with each other and probed against the experimental data.

In the present contribution we extend our investigations to the thermal and vibrational properties of the binary semiconductor ZnSe and ZnTe. ZnSe and ZnTe crystallize at normal pressure and temperature in the cubic zb structure. ZnSe is a wide-band-gap II-VI semiconductor with a gap of 2.8 eV while the band-gap of ZnTe is smaller and amounts to 2.2 eV. Because of its superior wide range infrared low-temperature optical transmittance ZnSe has gained technological importance for application as infrared transparent optical components, such as windows.
ZnTe can be easily doped and found applications in optoelectronics, e.g.  for blue light-emitting diodes and solar cells.

The vibrational and thermal properties of ZnSe and ZnTe have been the subject of a number of studies. Initially the comparison of experimental data such as the phonon dispersion branches of ZnSe and ZnTe  with inelastic neutron scattering (INS) (Ref. \onlinecite{Hennion1971,Vagelatos1974}) or the heat capacities with theoretical approaches remained on the semiempirical level.\cite{Plumelle1974,Talwar1981,Kushwaha2010}
These early investigations  have been extended recently to \textit{ab initio} LDA and GGA calculations using up-to-date implementations of the DFT method.  LDA results have been compared with the INS phonon energies and the heat capacities and found to be in good agreement.\cite{Corso1993,Petzke1997,Hamdi2006} Hamdi \textit{et al.} have also calculated  the structural  parameters of ZnSe and compared them with experimental and other available theoretical data.\cite{Hamdi2006} Recently, Tan \textit{et al.} have used DFT-GGA calculations to study the pressure induced phase transitions from the zb to the rocksalt structure.\cite{Tan2010} They also performed calculations of the phonon dispersion relations and heat capacities of ZnSe and ZnTe with the DFT-GGA CASTEP code.\cite{Castep} Unfortunately, the results of these calculations have not been compared  with the available experimental data in more detail.

Investigations of isotope substitution of the atomic masses of either Zn/(Se,Te) or simultaneously of both constituents simultaneously, on the properties of ZnX (X = Se, Te) appear to be rather scarce. To the best of our knowledge, the only available studies are a theoretical investigation of the temperature and mass dependence of the lattice constants, the temperature dependence of the linear thermal expansion coefficient and the mode Gr\"uneisen parameters upon isotopic substitution in ZnSe using perturbation theory in a density-functional framework.\cite{Debernardi1996}
Also, available is an experimental investigation of the Raman and excitonic spectra with different isotopic compositions.\cite{Goebel1998}
In Ref. \onlinecite{Goebel1998} an interesting dependence of the linewidth of the Raman spectra of ZnSe on isotope disorder was discovered.

We have recently demonstrated that changes of the specific heats due to the isotope substitution, even though sometimes very small, can be revealed by careful experiments even on very small crystalline samples.\cite{Cardona2010b} Especially, we have shown that the logarithmic derivatives with respect to the atomic masses $m_i$, $d\,ln\,\,(C_p/T^3)$/$d\,ln\,\,m_i$ are a very sensitive and  useful tool to  highlight small alterations of the specific heats since the logarithmic derivative eliminates, to a large extent, systematic experimental errors. In binary and multinary compounds the logarithmic derivatives particularly allowed us to reveal the influence of variations of the atomic masses on the phonon spectrum and to compare with the calculated phonon density of states (PDOS). We have also proposed and tested a sum rule which relates the sum of the logarithmic derivatives of $C_p/T^3$ with respect to the atomic masses with the logarithmic derivative with respect to temperature.\cite{Cardona2007}

The low-temperature heat capacities of ZnSe have been the subject of two preceding experimental studies carried out in the 1970s by Irwin and La Combe \cite{Irwin1974}  and by Birch.\cite{Birch1975}
However, the two previous heat capacity data sets just overlap in the temperature range of the $C_p/T^3$ maximum which for ZnSe occurs at about 20 K: in order to enable a meaningful comparison, e.g. with the logarithmic derivatives, and to test the sum rule a repetition of the measurements on a ZnSe sample with natural isotope composition  was found necessary.
We therefore carried out measurements of the heat capacities on a series of samples with various isotope combinations with special emphasis on the low-temperature regime.
For ZnTe heat capacity data are available only down to $\sim$15 K.\cite{Irwin1974,Demidenko1969,Gavrichev2002} The data from the different authors deviate significantly from each other, especially in the temperature regime where the maximum in $C_p/T^3$ is expected. We therefore repeated and extended the heat capacity measurement on ZnTe down to $\sim$ 2 K. Our measurements reveal clearly the maximum in   $C_p/T^3$ at $\sim$14 K. Additionally, our low temperature heat capacity data  enable us to extract very reliably the Debye temperatures for $T \rightarrow$ 0 K, $\Theta_{\rm Debye} (0)$ for ZnSe and ZnTe.

The comparison of our experimental data and literature data of the phonon dispersion obtained by INS experiments with  the calculations reveal significantly better agreement with the  LDA while the GGA data deviate noticeably from the experiments. The analysis of the total and the partial phonon densities of states and the comparison with that of ZnS (Ref. \onlinecite{Cardona2010a}) highlights the close similarities of the PDOS of  ZnX (X = S, Se, Te). It also reveals the compression of the frequency scales with increasing masses of the chalcogen atom. The  PDOSs projected on the Zn and the X atoms shows an interesting  trend on going from S to Te. While in ZnS the acoustical and optical phonons can be clearly attributed to vibrations of the Zn and S atoms, respectively, the situation is reversed for ZnTe. Here the optical branches emerge essentially from vibrations of the lighter Zn atoms. For ZnSe, with almost equal masses of the Zn and the Se atoms, the acoustical and optical phonon branches arise from combined vibrations of Zn and Se, respectively.

\section{Theoretical Details}

For the calculations reported here we have made use
of density functional theory\cite{Hohenberg1964,Kohn1965}
as implemented in the ABINIT package \cite{Abinitref}. Two different
exchange-correlation functionals (LDA, GGA) were tested but most of our
results are based on the LDA approach,
basically because it gives a much better agreement with experimental
data.

In the ABINIT package we have utilized a linear response
approach\cite{Baroni1987,Gonze1997A,Gonze1997B}  together with an
iterative minimization norm-conserving pseudopotential plane-wave
method.\cite{Gonze2002}
These pseudopotentials are single projector, ordinary norm conserving,
based on the Troullier-Martins
method.\cite{Hamann1979,Troullier1991,Payne1992} 12 and 6 "valence"
electrons were used  for Zn and Se, respectively. A 40 Ry cutoff was set
for the plane wave expansion and an 8 $\times$ 8 $\times$ 8
\textbf{k}-point grid regularly shifted along four different
directions, with a total of 60 \textbf{k}-points for the ground state calculation
and 2048
\textbf{k}-points for the vibrational response part.
The calculations converged well: increasing the plane-wave cutoff to 50 Ry
and  the \textbf{k}-points mesh to 10 $\times$ 10 $\times$ 10 yielded, on
average, a change of  less then two percent in the phonon frequencies. For
the phonon frequencies we have used a \textbf{q}-mesh of
12 $\times$ 12 $\times$ 12, to guarantee a good coverage of the dispersion
relations. The dynamical matrices were obtained from perturbation
theory\cite{Gonze1997A,Gonze1997B} and a Fourier interpolation was
employed in order to increase the mesh sampling and to improve the
description of quantities such as vibrational density of states and heat
capacities.

\section{Experimental}
zb-ZnSe and zb-ZnTe crystals with different isotopic compositions:
$^{\rm nat}$Zn $^{\rm nat}$Se/Te i.e., Zn and Se/Te with natural isotope composition,  $^{64}$Zn $^{76}$Se,  $^{64}$Zn $^{80}$Se, $^{68}$Zn $^{76}$Se, and $^{68}$Zn $^{80}$Se, $^{68}$Zn $^{\rm nat}$Te, and $^{\rm nat}$Zn $^{130}$Te,
were grown by vapor phase transport as described in detail previously.\cite{Lauck1999}.  Larger zb-ZnTe crystals with the natural isotope composition were obtained by a modified Piper-Polich method in a semi-open quartz-glass ampoule.\cite{Lauck1986a,Lauck1986b}
The heat capacities
of crystalline pieces of these samples,  typically of $\sim$20 - 100 mg weight,  were measured between 2 and
280 K with a physical property measurement system (Quantum
Design, San Diego, CA) as described in detail in Ref. \cite{Serrano2006}

\section{Lattice Dynamics}

\begin{figure}[ht]
\includegraphics[width=7.5cm ]{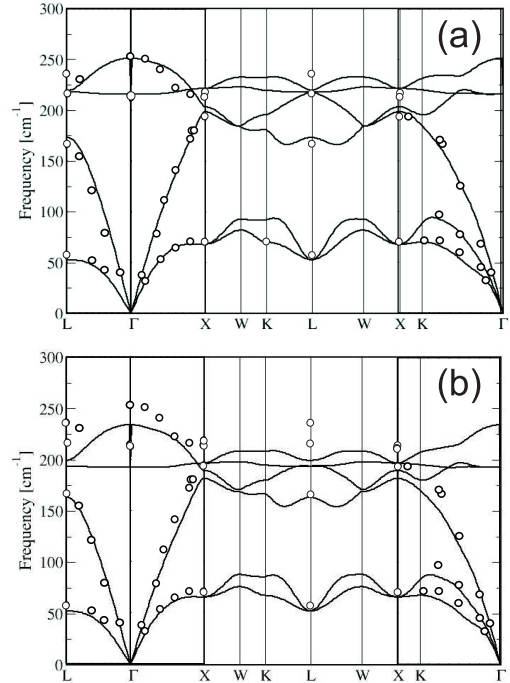}
\caption{Phonon dispersion relations of ZnSe as obtained by DFT-LDA (a) and DFT-GGA (b) calculations. The circles represent the inelastic neutron scattering data measured by Hennion \textit{et al.}.\cite{Hennion1971} } \label{Fig1}
\end{figure}

In Figure \ref{Fig1}(a) and (b) and  we display the calculated phonon dispersion relations of ZnSe  along selected directions of the Brillouin zone (BZ).
For comparison we have  plotted our LDA and GGA results and the
INS data reported by Hennion \textit{et al.}\cite{Hennion1971}
The LDA calculations  are overall in good agreement with the experimental findings, as has also been  observed by Dal Corso \textit{et al.} \cite{Corso1993}  The results of the GGA calculation deviate markedly, especially for the optical branches.

\begin{figure}[ht]
\includegraphics[width=7.5cm ]{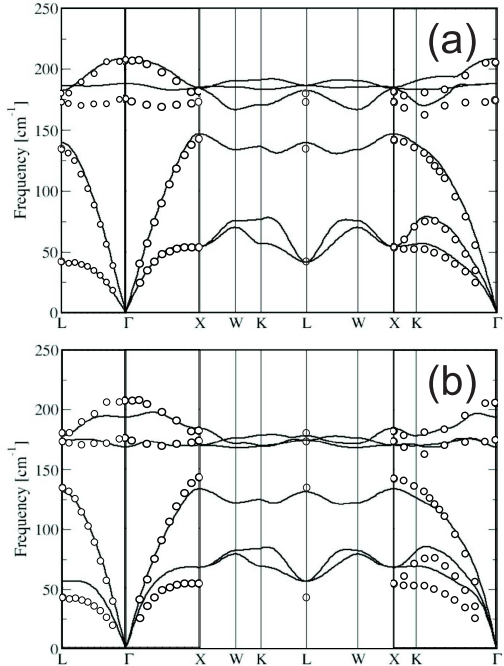}
\caption{Phonon dispersion relations of ZnTe as obtained by DFT-LDA (a) and DFT-GGA (b) calculations. The circles represent the inelastic neutron scattering data measured by Vagelatos \textit{et al.}.\cite{Vagelatos1974} } \label{Fig1b}
\end{figure}

Figure \ref{Fig1b}(a) and (b) show the comparison of the LDA and GGA calculations for ZnTe with the INS data collected by Vagelatos \textit{et al.}.\cite{Vagelatos1974} As already observed by Dal Corso \textit{et al.} the agreement of the experiment with the DFT-LDA calculations is excellent, especially for the acoustical branches. The longitudinal optical branches are equally well reproduced by the calculations, while the transverse optical branches fall short by about 5\%.
The DFT-GGA results agree well with the optical branches while the energies of transverse acoustical branches lie systematically below the experimental data.

\begin{figure}[ht]
\includegraphics[width=7.9cm ]{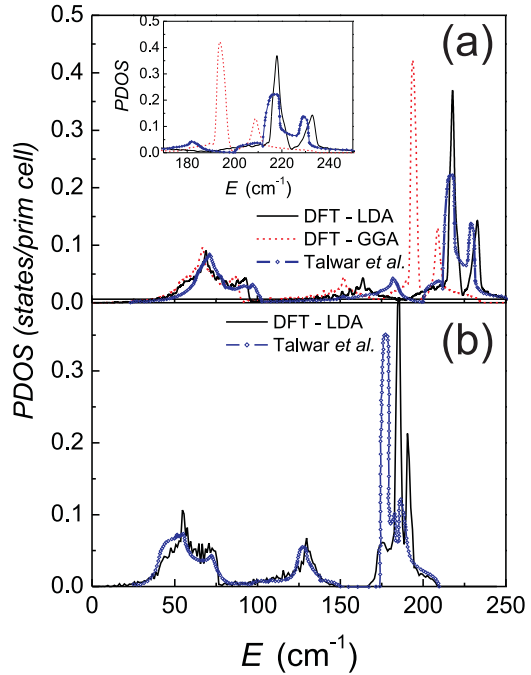}
\caption{One-phonon density of states (PDOS) of (a) ZnSe and (b) ZnTe as obtained by DFT-LDA  and DFT-GGA  calculations. The diamonds represent data by Talwar \textit{et al.}.\cite{Talwar1981} The inset displays the high energy section in an enlarged scale.} \label{Fig2}
\end{figure}

\begin{figure}[ht]
\includegraphics[width=7.9cm ]{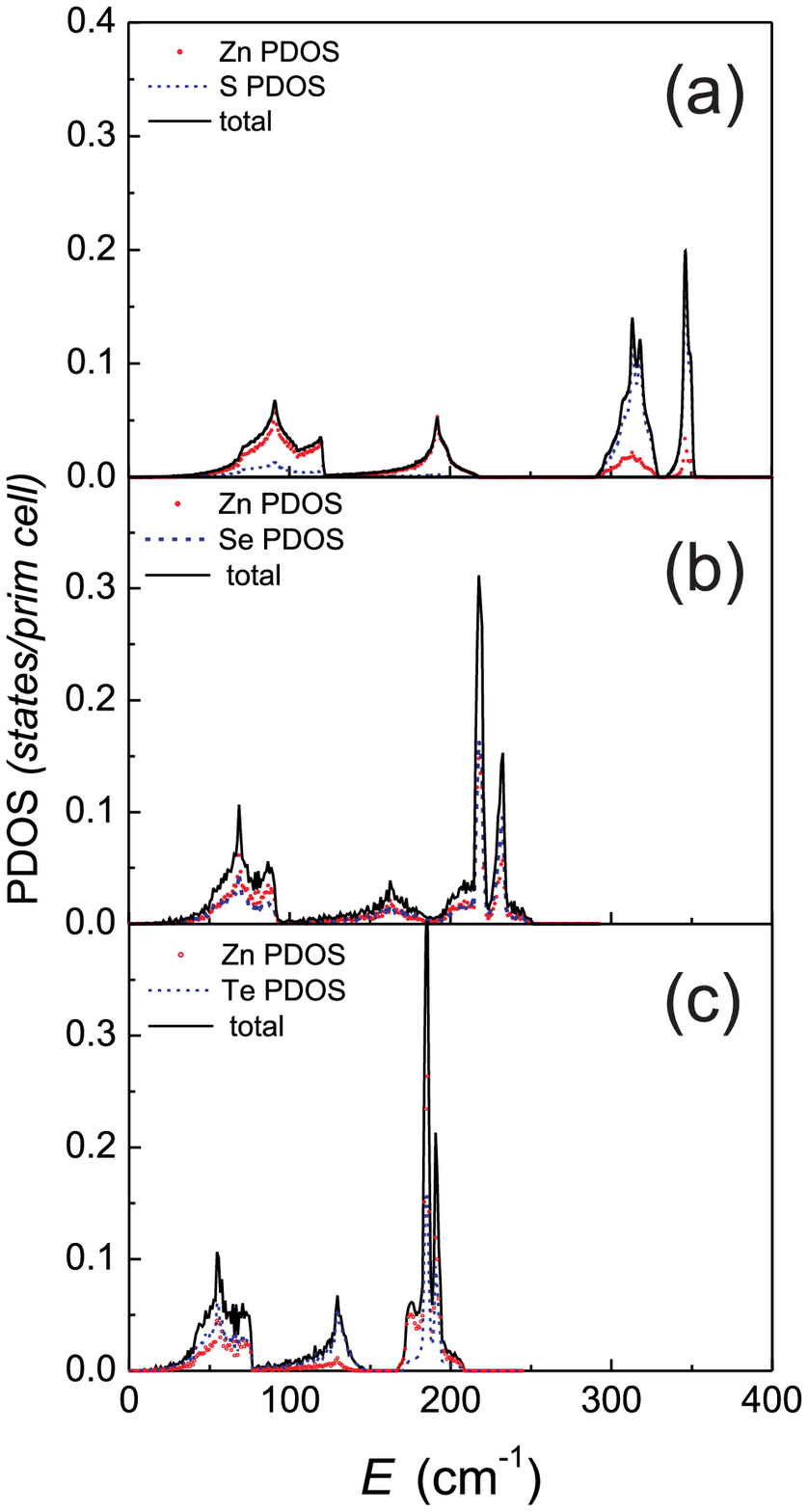}
\caption{Comparison of the one-phonon density of states (PDOS) of (a) ZnS (Ref. \onlinecite{Cardona2010a}), (b) ZnSe and (c) ZnTe as obtained by DFT-LDA   calculations. The decomposition into contributions from Zn and from the chalcogen is also given.\cite{Talwar1981} } \label{Fig2b}
\end{figure}

The total  PDOS and the PDOS projected on the two constituting elements as obtained by the LDA calculations, are displayed in Fig. \ref{Fig2}. For comparison we also show the PDOS of ZnSe and ZnTe derived by Talwar \textit{et al.} from a rigid-ion-model.\cite{Talwar1981}
In Figure \ref{Fig2b} we also added the PDOS of ZnS as published by us recently.\cite{Cardona2010a}
Very similar to the PDOS of  ZnS the PDOSs of ZnSe and ZnTe exhibit two dominating features, one between  50 and 100~cm$^{-1}$, arising essentially from acoustic phonon branches, and    a sharp double peak  at higher energies corresponding to the the optical phonons. The energies of the latter peaks decrease markedly as one substitutes S by Se and Te.
These two main peak groups are separated by a gap in which a small mid-gap peak is located.
While the low-energy peak in the PDOS of ZnSe is largely independent of the computational approach  the mid-gap feature and the sharp optical phonon double-peak obtained from the GGA calculations are  down-shifted, the latter by about 10\%  as compared to the LDA result. Except for the mid-gap peak the LDA PDOS is generally in good agreement with results of the rigid-ion-model used by Talwar \textit{et al.}.\cite{Talwar1981}

It is of interest to compare the projected PDOS of the three systems ZnX (X =S, Se, Te) (see Fig. \ref{Fig2b}). For the ZnS the acoustical and the high energy optical phonon branches arise essentially from vibrations of Zn and S, respectively. Due to the reversal of the  masses of the two constituents in ZnTe (Zn: $\sim$65 a.m.u.; Te: $\sim$128 a.m.u) the situation is reversed for ZnTe. There the acoustical phonons arise from Te vibrations and the optical ones from vibrations of the lighter Zn atoms. Due to the very similar atomic masses of Zn and Se   acoustical and optical phonon branches for ZnSe comprise mixed Zn and Se vibrations.

\section{Heat Capacity}

By using the  calculated PDOS($\omega$) we obtained the free energy, $F(T)$ and the heat capacity (at constant volume), $C_v(T)$ according to\cite{CvCp}

\begin{eqnarray}
    \label{Eq1}
    F(T) &=& -\int\limits_0^\infty \left\{\frac{\hbar\omega}{2}+k_B T\ln[2n_B(\omega)]\right\}\rho(\omega)d\omega,\\
    \label{eq:cv}
    C_{v}(T) &=& -T\,\left(\frac{\partial^2 F}{\partial T^2} \right)_{\rm V},
\end{eqnarray}

where $k_B$ is the Boltzmann constant and $n_B$ the Bose-Einstein
factor. The cut-off in  PDOS$(\omega)$ at the highest phonon frequency, defines the upper limit of integration.

\begin{figure}[ht]
\includegraphics[width=8.1cm ]{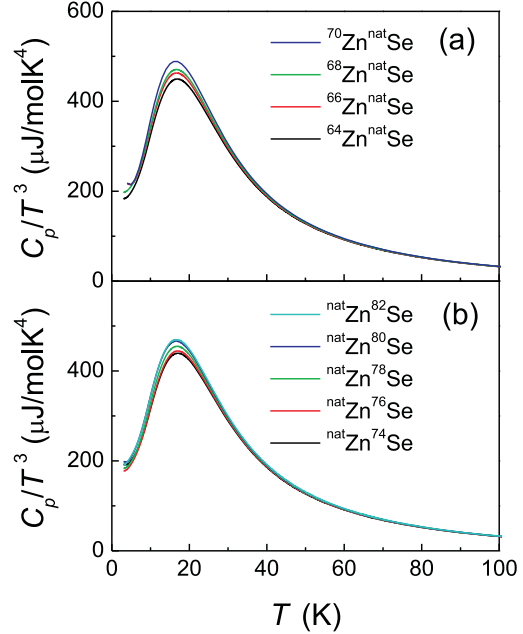}
\caption{Calculated low-temperature heat capacity of ZnSe (ABINIT DFT-LDA) assuming different masses for the Zn and the Se isotopes, as given in the inserts (increasing isotope mass from bottom to top).}  \label{Fig3}
\end{figure}

\begin{figure}[ht]
\includegraphics[width=8.3cm ]{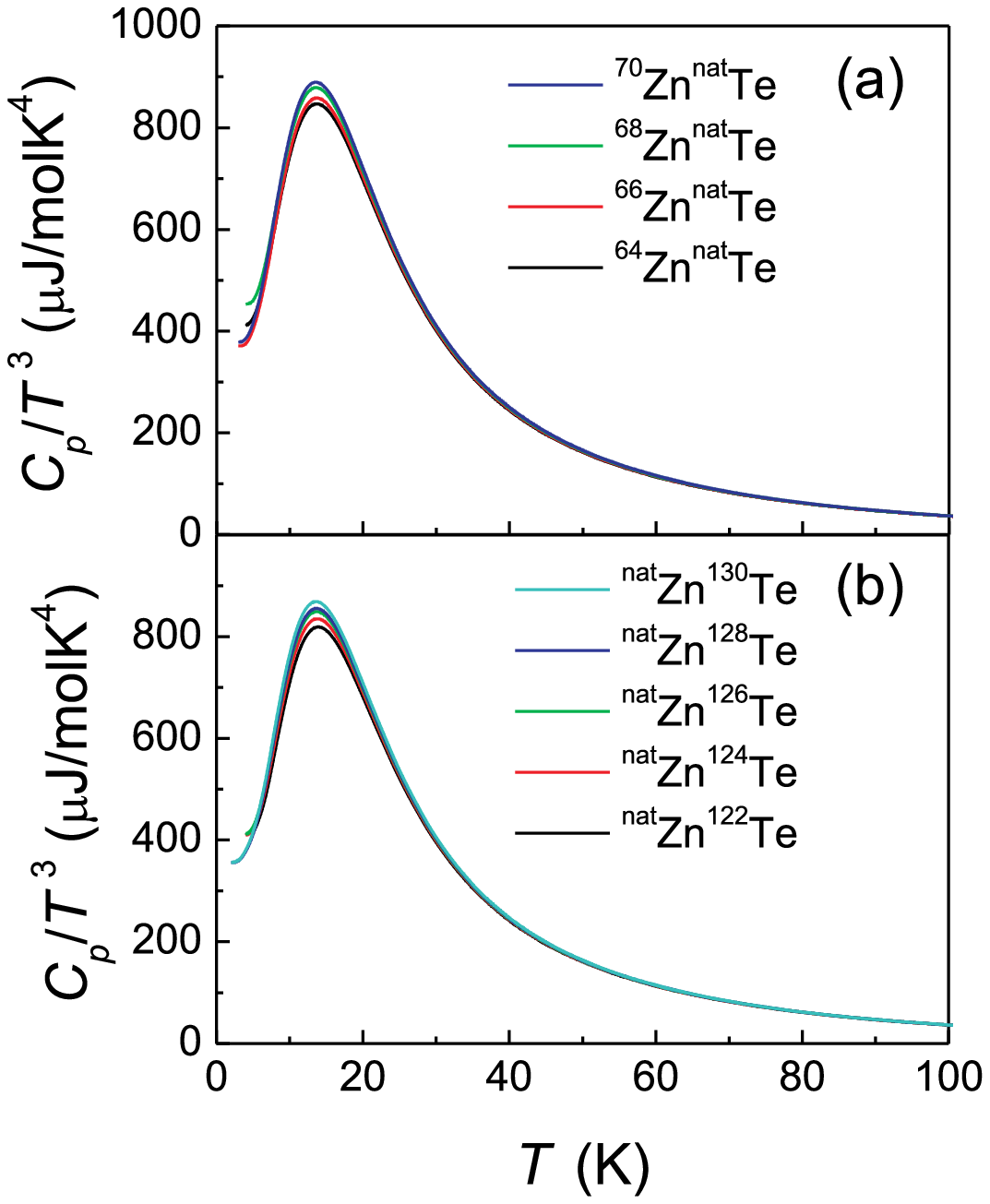}
\caption{Calculated low-temperature heat capacity of ZnTe (ABINIT DFT-LDA) assuming different masses for the Zn and the Te isotopes, as given in the inserts (increasing isotope mass from bottom to top).}  \label{Fig3b}
\end{figure}

Figure \ref{Fig3} and \ref{Fig3b} display the calculated heat capacities, $C_p \approx  C_v(T)$ of ZnSe and ZnTe for different isotope mass compositions in the $C_p/T^3(T)$ representation which is chosen to emphasize the broad peak at low temperature. For ZnSe the maximum is found  at 17.2 K and for ZnTe it moves down to 13.5 K.
The broad peak in the  $C_p/T^3(T)$ representation of the heat capacity has been demonstrated to arise from regions of low dispersion of zone-boundary acoustical phonons which also give rise to the low-energy peaks in the PDOS.\cite{Cardona2007}
They appear at 68.6 cm$^{-1}$ and 55.1 cm$^{-1}$ for ZnSe and ZnTe, respectively. We again find a ratio  $\sim$ 6 for the energies of the acoustical phonon peaks and the peak temperature in the $C_p/T^3(T)$ representation, close to what has been typically observed for several semiconductors before.\cite{Cardona2010b}

\begin{figure}[ht]
\includegraphics[width=8.1cm ]{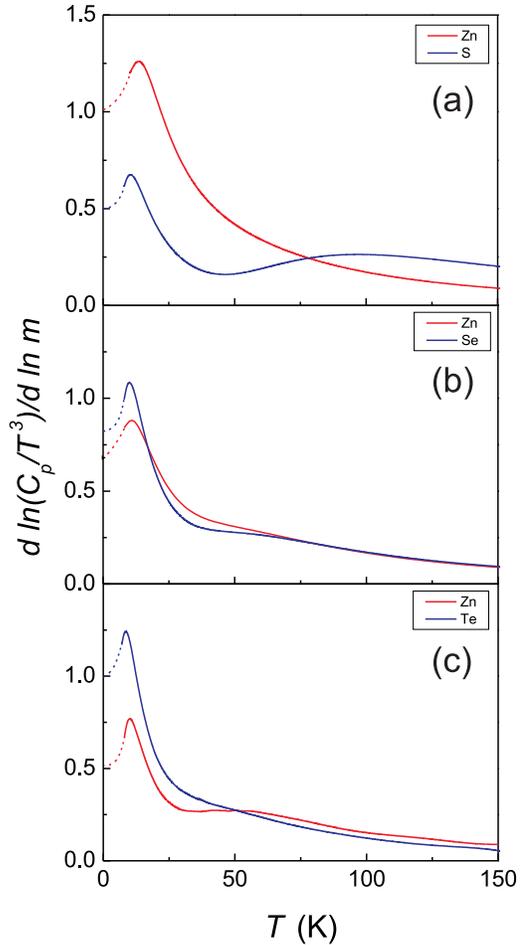}
\caption{(Color online) Logarithmic derivatives vs the masses of the constituents, Zn (solid red) and X (X = S, Se, Te; from top to bottom, see inset) (solid blue) calculated from the LDA heat capacity results displayed in Figs. \ref{Fig3} and \ref{Fig3b}. The ZnS data (a) were taken from Ref. \onlinecite{Cardona2010a}. The dashed part of the calculated curves have been
extrapolated by free hand so as to agree with the value of Eqs. (\ref{Eq2a}) and (\ref{Eq2b}) at
$T$ = 0.}  \label{Fig3c}
\end{figure}

The peak value of $C_p/T^3(\sim 13.5 K)$ for ZnTe is by almost a factor of two larger than that of ZnSe at 17.2 K, reflecting the lattice softening when Se is replaced by Te.
Isotope mass increases lead to small growth of the heat capacities, which become clearly visible as slight changes of the maximum value of the $C_p/T^3(T)$ peak. Below we will discuss the logarithmic derivatives of $C_p/T^3(T)$ which reveal that the variation of the isotope masses changes the heat capacities in different temperature regimes.
Figure \ref{Fig3c} displays the logarithmic derivatives,
$d\,ln\,\,(C_p/T^3)$/$d\,ln\,\,m_i$,
with respect to the isotope mass, $m_i$,  of the two constituting elements, Zn and either Se or Te, respectively.  For completeness we have also included the  analog graphs for ZnS.\cite{Cardona2010a} For the latter there is a clear differentiation between the derivative vs the masses of Zn and S. Similar sharp peaks at low temperature, with that of S significantly smaller in magnitude than that of the Zn derivative, have been attributed to a considerable S component in the TA vibrations which have mainly Zn character. The characteristic S contribution due to high energy S vibrations shows up as a broad peak centered near 100 K.\cite{Cardona2010a}
Similar derivatives are obtained for ZnSe and ZnTe. However,  depending on the relative masses of the constituents Zn and Se/Te the magnitude and the maximum temperatures of the peaks shift. For ZnSe the low-temperature peaks in the derivative are of equal magnitude and a pronounced high-temperature peak is missing. This finding reflects the mixing of Zn- and Se-related vibrations due to the rather similar atomic masses of Zn and Se. For ZnTe the situation is reversed. Here the pronounced peak at low temperature is associated to Te vibrations, while the vibrations of the lighter Zn atoms give rise to a shallow broad peak centered at $\sim$ 70 K. Again, as in ZnS, there is some hybridization of Zn with Te vibrations and a noticeable Zn component in the TA vibrations which now have mainly Te character.

In previous work we have demonstrated that for low temperatures, $T \rightarrow$ 0, the logarithmic derivatives are related to the ratios of the atomic mass to the molar mass according to,\cite{Serrano2006,Cardona2010a,Romero2011}

\begin{equation}
    \frac{d\ln(C_{p}/T^3)}{d\ln m_{\rm Zn}} = \frac{3}{2}\,\frac{m_{\rm Zn}}{m_{\rm Zn}+m_{\rm X}} = 0.68 \rm{(Se)}; = 0.51 \rm{(Te)}\label{Eq2a}
\end{equation}

\begin{equation}
    \frac{d\ln(C_{p}/T^3)}{d\ln m_{\rm X }} = \frac{3}{2}\,\frac{m_{\rm X }}{m_{\rm Zn}+m_{\rm X}} = 0.82 \rm{(Se)}; = 0.99 \rm{(Te)}.
    \label{Eq2b}
\end{equation}

The dashed lines at low temperatures in Figure \ref{Fig3c} represent a free-hand extrapolation for $T \rightarrow$ 0 so as to agree with these values.

\begin{figure}[ht]
\includegraphics[width=8.2cm ]{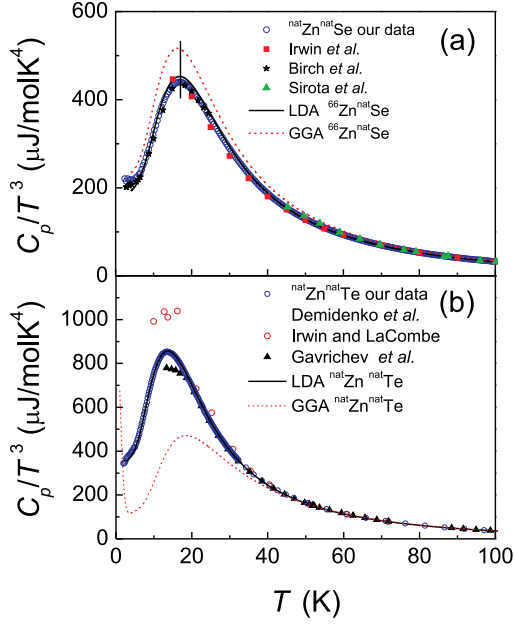}
\caption{Our experimental results and literature data of the low-temperature heat capacities  of (a) ZnSe and (b) ZnTe compared with the ABINIT DFT-LDA and GGA results, (black) solid and (red) dashed lines, respectively. For the references of the literature data see text.}  \label{Fig4}
\end{figure}

Our low temperature heat capacities of ZnSe and ZnTe collected on samples with the natural isotope composition are summarized and compared with the results of the DFT-LDA and GGA calculations in Figure \ref{Fig4}. The $C_p/T^3$ representation reveals low-temperature peaks at 17.0 K and at 13.5 K, for ZnSe and ZnTe, respectively. For ZnSe the data by Irwin \textit{et al.}\cite{Irwin1974} and Birch\cite{Birch1975} are in good agreement with our more complete data.
Especially, for ZnTe we could resolve the diverging low-temperature heat capacity results obtained before by Gavrichev \textit{et al.}\cite{Gavrichev2002} and Irwin \textit{et al.}\cite{Irwin1974} and extend the temperature range down to 2 K. This was found necessary in order to clearly establish the position and magnitude of the $C_p/T^3$ maximum and to enable the calculation of the logarithmic derivatives, discussed below.

The Debye temperatures, $\theta_{Debye}(0)$, for $T \rightarrow$ 0 K are obtained from  the experimental data by fitting a power law
\begin{equation}
C_p/T = \beta T^2 + \delta T^4
\label{EqDebye}
\end{equation}

\noindent to the low-temperature range 2 K $\leq T \leq$ 10 K.

\noindent For ZnSe we obtained

\begin{equation*}
\theta_{\rm Debye}(0) = 289 (2) \rm{K} \,\,\, \rm{for\,\,\,ZnSe},
\end{equation*}

which is by about 6\% larger than the value  reported by Birch \cite{Birch1975} and somewhat closer to the result (278.5 K) derived from the elastic constants by Lee.\cite{Lee1970}

For ZnTe we found

\begin{equation*}
\theta_{\rm Debye}(0) = 230 (2) \rm{K}\,\,\, \rm{for\,\,\,ZnTe}
\end{equation*}

\noindent which is close to the value from elastic constants measurements 225.3 K\cite{Lee1970} but significantly larger than the values
estimated  from x-ray measurements and quoted in review articles (182 K - 210 K).\cite{Blattner1972,Landolt1999}

We finally note that the ratio of the Debye-temperatures, 1.26,  is very close to the square-root of the inverse atomic mass ratio of the chalcogenide atoms,

\begin{equation*}
\theta_{\rm Debye}(0)_{\rm ZnSe}/\theta_{\rm Debye}(0)_{\rm ZnTe} \approx \sqrt{m_{\rm Te}/m_{\rm Se}} = 1.27.
\end{equation*}

As already observed for the phonon dispersion,  the agrement of the heat capacities of ZnSe and ZnTe with the LDA calculations is better than with the  GGA results. Especially in the temperature regime of the $C_p/T^3$ maximum overestimates/underestimates  the experimental data by about 20\%/50\% for ZnSe and ZnTe, respectively.

\begin{figure}[ht]
\includegraphics[width=8.2cm ]{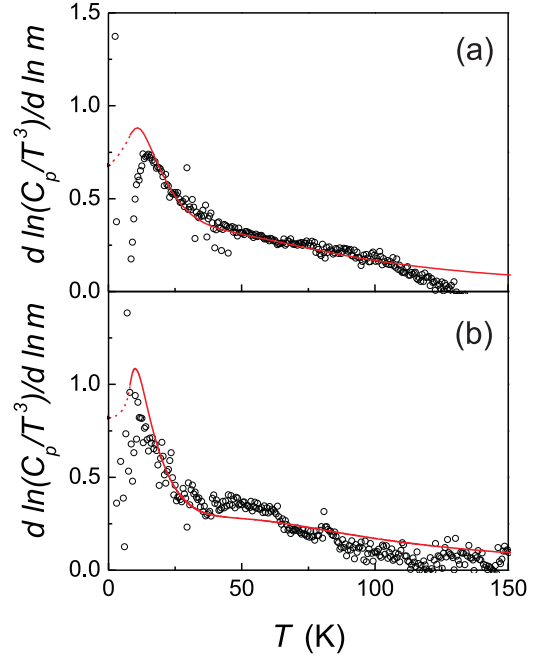}
\caption{Logarithmic derivative of $C_p/T^3$ of ZnSe vs the isotope masses of (a) Zn and (b) Se. (o) experimental results, (red) solid line LDA calculations. The (red) dashed lines are free-hand extrapolations (see Fig. \ref{Fig3c}).}  \label{Fig5}
\end{figure}

\begin{figure}[ht]
\includegraphics[width=8.2cm ]{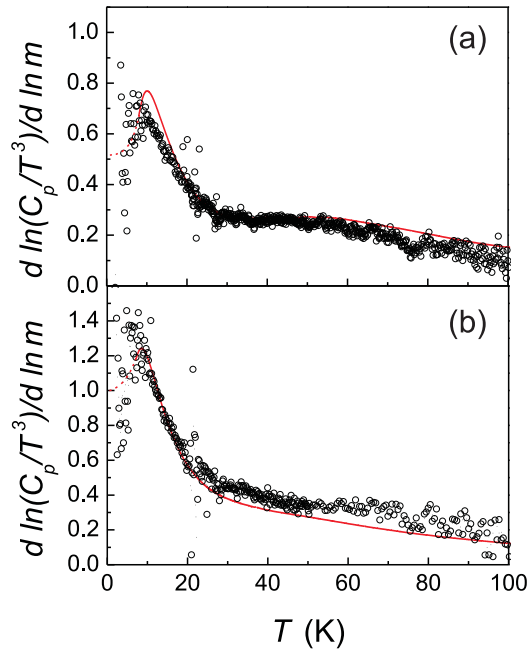}
\caption{Logarithmic derivative of $C_p/T^3$ of ZnTe vs the isotope masses of (a) Zn and (b) Te. (o) experimental results, (red) solid line LDA calculations. The (red) dashed lines are free-hand extrapolations (see Fig. \ref{Fig3c}).}  \label{Fig7}
\end{figure}

The logarithmic derivatives vs the mass of the isotopes of Zn and Se are summarized in Figure \ref{Fig5}. Those for Zn and Te are displayed in Figure \ref{Fig7}. The agreement with the LDA calculations in magnitude and position of the low temperature peaks is very satisfactory. For ZnTe, the logarithmic derivative with respect to the isotope mass of the Zn atoms clearly exhibits the broad maximum centered at $\sim$50 K due to phonons involving low-energy Zn vibrations.

In previous work, on monatomic crystals we derived a connection
between the logarithmic derivative of $C_p/T^3$ versus $T$
and the corresponding derivative vs the isotopic mass.\cite{Gibin2005}
For binary and ternary
materials, a similar connection was shown to hold provided
one adds the two or three logarithmic derivatives with respect to each of
the isotope masses.\cite{Cardona2007,Romero2011}

\begin{figure}[ht]
\includegraphics[width=8.2cm ]{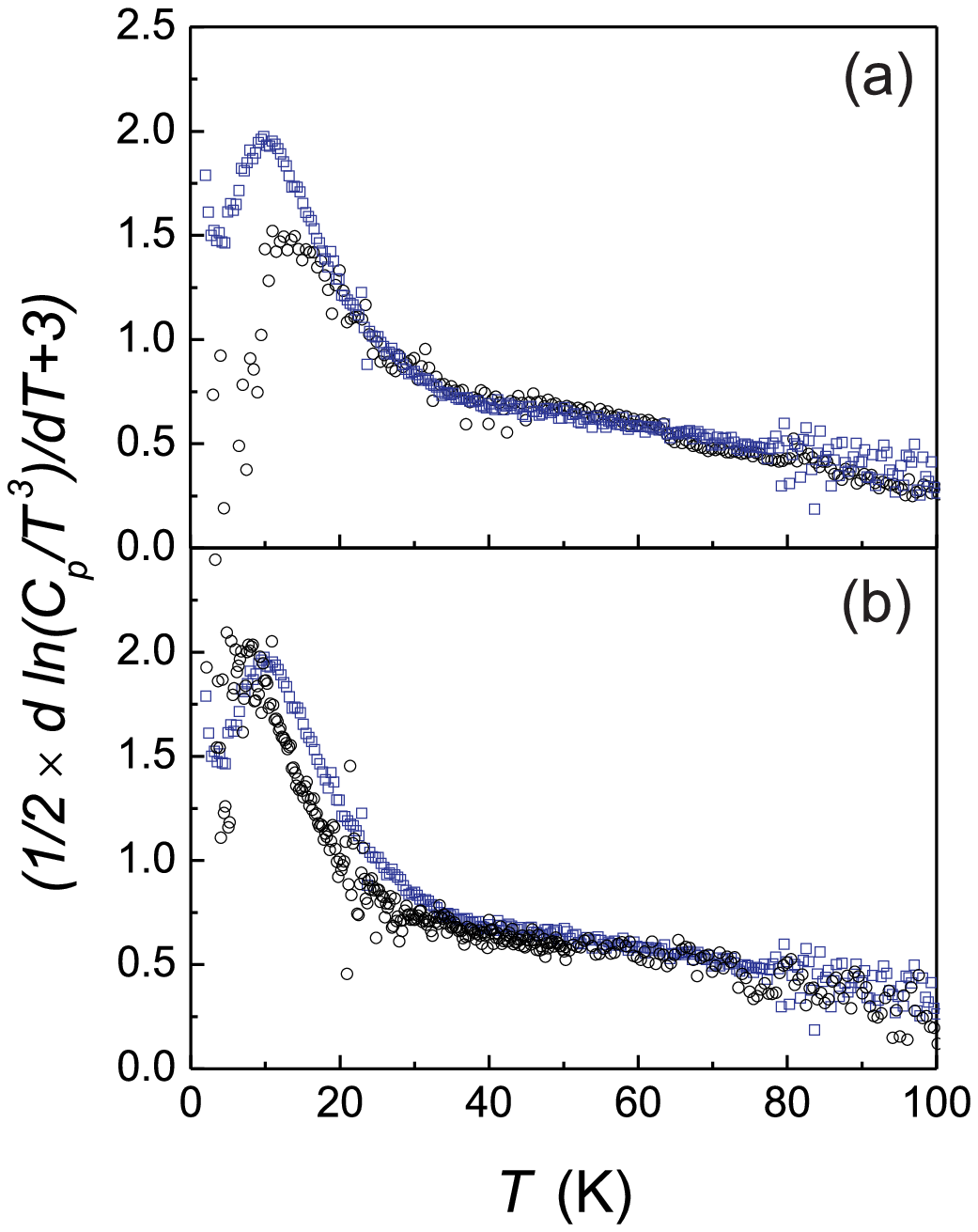}
\caption{Comparison and sum rule (Eq. (\ref{Eq3})) for (a) ZnSe and (b) ZnTe. The circles and the squares represent the r.h.s and l.h.s. of Eq. (\ref{Eq3}), respectively.}  \label{Fig11}
\end{figure}

Accordingly, for the relation of the temperature and isotope mass derivatives in a two component system we obtain the expression \cite{Romero2008}


\begin{equation}
  \frac{1}{2}\,\,(3 + \frac{d
\ln (C_{p}(T)/T^3)}{d \ln T})=\sum_{i=1}^2 \frac{d \ln (C_{p}(T)/T^3)}{d \ln m_{\rm i}}, \label{Eq3}
\end{equation}

where $m_i$ (i =1, 2) are the masses of the two constituting elements, i.e. Zn and X = Se and Te, respectively.
As demonstrated in Figure \ref{Fig11},  the  relation given in Eq. (\ref{Eq3}) is nicely fulfilled for ZnSe and ZnTe, thus lending further support for the proposed sum rule.

\section{Conclusions}\label{SectionConclus}

Up-to-date \textit{ab initio} electronic structure  calculations have now matured and became a powerful tool to study the thermodynamic properties of crystals. Here we demonstrate this for the binary zinc chalcogenides, ZnSe and ZnTe. Both materials are technologically important semiconductors used for optical applications. We have used DFT-LDA and GGA codes to calculate the phonon dispersions and the heat capacities down to lowest temperatures. In the experimental part we extend the available experimental basis and comprehensively determine the heat capacities of ZnSe and ZnTe over the temperature range from 2 K to room temperature and have compared the results with the calculations. Generally, better agreement of the experimental data is found with the DFT-LDA results. Additionally, we measured the heat capacities of isotope substituted samples in order to vary the atomic masses of Zn and Se and Te and to probe the effect of selective isotope substitution on the phonon spectra and the thermodynamic properties. These data are compared with the results of the calculations and found to be in very good agreement.  A comparison of ZnS, ZnSe and ZnTe instructively reveals the variation of the phonon spectrum with the variation of the atom mass of chalcogenide atoms.

\begin{acknowledgments}
AHR has been supported by CONACYT Mexico under projects J-152153-F, Binational
Collaboration FNRS-Belgium-CONACYT and PPPROALMEX-DAAD-CONACYT.
We are indebted to N.N.  for a critical reading of the manuscript.

\end{acknowledgments}


\begin{thebibliography}{99}





\bibitem{Cardona2009}
M. Cardona, R. K. Kremer,
R. Lauck, G. Siegle, A. Mu\~{n}oz, and A. H. Romero, Phys. Rev. B \textbf{80}, 195204 (2009).

\bibitem{Romero2008}
A. H. Romero,
M. Cardona, R. K. Kremer, R. Lauck,  G. Siegle,  J. Serrano,
and  X. C. Gonze, Phys. Rev. B \textbf{78}, 224302 (2008).


\bibitem{Cardona2010b}
M. Cardona, R. K. Kremer, G. Siegle,
A. Mu\~{n}oz, A. H. Romero, and  M. Schmidt, Phys. Rev. B \textbf{82}, 085210 (2010).

\bibitem{Cardona2010a}
M. Cardona, R. K. Kremer, R. Lauck, G. Siegle, A. Mu\~{n}oz, and A. H. Romero, and A. Schindler,
Phys. Rev. B \textbf{81} 075207 (2010).



\bibitem{Romero2011}
A. H. Romero,
M. Cardona, R. K. Kremer , R. Lauck, G. Siegle,  C. Hoch, A. Mu\~{n}oz, and
A. Schindler, Phys. Rev B \textbf{83}, 195208 (2011).


\bibitem{Kremer2005}
R. K. Kremer, M. Cardona, E. Schmitt, J. Blumm, S.K. Estreicher, M. Sanati, M. Bockowski, I. Grzegory,
T. Suski, and A. Jezowski, Phys. Rev. B \textbf{72}, 075209 (2005).

\bibitem{Gibin2005}
A. Gibin, G.G. Devyatykh, A.V. Gusev, R.K. Kremer, M. Cardona, and
H.-J. Pohl,
 Sol. State Comm. \textbf{133}, 569 (2005).

\bibitem{Diaz2007}
L. E. Diaz-Sanchez, A. H. Romero, M. Cardona, R. K. Kremer, and X. Gonze, Phys. Rev. Lett. \textbf{99}, 165504 (2007).



\bibitem{Serrano2008}
J. Serrano, R. K. Kremer, M. Cardona, G. Siegle, L. E. Diaz-Sanchez, and A. H. Romero,
Phys. Rev. B \textbf{77}, 054303 (2008).






\bibitem{Hennion1971}
B. Hennion, F. Moussa, G. Pepy, and K. Kunc, Phys. Lett. \textbf{36A}, 376 (1971).


\bibitem{Vagelatos1974}
N. Vagelatos, D. Wehe, and J. S. King, J. Chem. Phys. \textbf{60}, 3613 (1974).








\bibitem{Plumelle1974}
P. Plumelle and  M. Vandevyer,
phys. stat. sol. (b) \textbf{73}, 271 (1976).

\bibitem{Talwar1981}
D. N. Talwar, M. Vandevyver, K. Kunc, and M. Zigone,
Phys. Rev B \textbf{24}, 741 (1981).
%

\bibitem{Kushwaha2010}
A.	K. Kushwaha, Physica B \textbf{405}, 1638 (2010).



\bibitem{Corso1993}
A. Dal Corso, S. Baroni, R. Resta, and S. de Gironcoli, Phys. Rev  B \textbf{47}, 3588 (1993).


\bibitem{Petzke1997}
K. Petzke, C.  Schrepel, U. Scherz,
Z. Phys. Chem. \textbf{201}, 317 (1997).




\bibitem{Hamdi2006}
I. Hamdi, M. Aouissi, A. Qteish, and N. Meskini,
Phys. Rev. B \textbf{73}, 174114 (2006).








\bibitem{Tan2010}
Tan, Jia-Jin, Ji, Guang-Fu, Chen, Xiang-Rong, and Guo, Quing-Quan, Commun. Theor. Phys. (Beijing, China) \textbf{53}, 1160 (2010).


\bibitem{Castep}
for reference see the CASTEP website at http://www.castep.org/.

\bibitem{Debernardi1996}
A.	Debernardi and M. Cardona, Phys. Rev. B \textbf{54}, 11305 (1996).

\bibitem{Goebel1998}
A.	G\"obel, T. Ruf, J. M. Zhang, R. Lauck, and M. Cardona,
Phys. Rev B \textbf{59}, 2749 (1999).



\bibitem{Cardona2007}
M. Cardona, R. K. Kremer, R. Lauck,  G. Siegle,
J. Serrano, and A. H. Romero, Phys. Rev. B \textbf{76}, 075211 (2007).


\bibitem{Irwin1974}
J. C. Irwin and J. LaCombe, J. Appl. Phys \textbf{45}, 567  (1974).



\bibitem{Birch1975}
J. A. Birch J. Phys C: Solid State Phys.  \textbf{8}, 2043 (1975).




\bibitem{Demidenko1969}
A.F. Demidenko and A.K. Maltsev, Izv. Akad. Nauk SSSR Neorg. Mater. \textbf{5}, 158 (1969).

\bibitem{Gavrichev2002}
K.S. Gavrichev, G.A. Sharpataya, V.N. Guskov,
J.H. Greenberg, T. Feltgen, M. Fiederle, and K.W. Benz,
phys. stat. sol. (b) \textbf{229}, 133 (2002).



















\bibitem{Hohenberg1964}
 P. Hohenberg and W. Kohn, Phys. Rev. \textbf{136}, B864 (1964).



\bibitem{Kohn1965}
W. Kohn and L. J. Sham, Phys. Rev. \textbf{140}, A1133  (1965).


\bibitem{Abinitref}
X. Gonze, B. Amadon, P.-M. Anglade, J.-M. Beuken, F. Bottin, P.
Boulanger, F. Bruneval,
D. Caliste, R. Caracas, M. Cote, T. Deutsch, L. Genovese, Ph. Ghosez, M.
Giantomassi
S. Goedecker, D.R. Hamann, P. Hermet, F. Jollet, G. Jomard, S. Leroux, M.
Mancini, S. Mazevet,
M.J.T. Oliveira, G. Onida, Y. Pouillon, T. Rangel, G.-M. Rignanese, D.
Sangalli, R. Shaltaf,
M. Torrent, M.J. Verstraete, G. Zerah, and J.W. Zwanziger,
Comp. Phys. Comm. \textbf{180}, 2582 (2009);
ABINIT is a common project of the Universite Catholique de Louvain, Corning Incorporated, and other contributors http://www.pcpm.ucl.ac.be/ABINIT.




\bibitem{Baroni1987}
S. Baroni, P. Giannozzi, and A. Testa, Phys. Rev. Lett. \textbf{58}, 1861 (1987).



\bibitem{Gonze1997A}
X. Gonze, Phys. Rev. B \textbf{55}, 10337  (1997).

\bibitem{Gonze1997B}
X. Gonze and C. Lee, Phys. Rev. B \textbf{55}, 10355  (1997).



\bibitem{Gonze2002}
X. Gonze, J. M. Beuken, R. Caracas, F. Detraux, M. Fuchs, G.  M. Rignanese, L. Sindic, M. Verstraete, G. Zerah, and F. Jollet, Comput. Mater. Sci. \textbf{25}, 478  (2002).





\bibitem{Hamann1979}
D. R. Hamann, M. Schluter and C. Chiang, Phys. Rev. Lett. \textbf{43}, 1494 (1979).

\bibitem{Troullier1991}
N. Troullier and Jos\'{e} Lu\'{\i}s Martins, Phys. Rev. B \textbf{43}, 1993 (1991).

\bibitem{Payne1992}
M. C. Payne, M. P. Teter, D. C. Allan, T. A. Arias, and J. D. Joannopoulos, Rev. Mod. Phys. \textbf{64}, 1045 (1992).






\bibitem{Lauck1999}
R. Lauck and E. Sch\"onherr, J. Cryst. Growth \textbf{197}, 513 (1999).

\bibitem{Lauck1986a}
R. Lauck and G. M\"uller-Vogt, J. Cryst. Growth \textbf{74}, 513 (1986).

\bibitem{Lauck1986b}
R. Lauck, G. M\"uller-Vogt, and W. Wendl, J. Cryst. Growth \textbf{74}, 520 (1986).

\bibitem{Serrano2006}
J. Serrano, R. K. Kremer, M. Cardona, G. Siegle, A. H. Romero, and R. Lauck, Phys. Rev. B \textbf{73}, 094303 (2006).

\bibitem{CvCp}
As discussed in
Ref.~\onlinecite{Kremer2005}, the difference between the experimentally accessible heat capacity at constant pressure,$C_{p}$, and
$C_{v}$ is significant only at high temperatures, thus negligible in
low temperature range considered here.


\bibitem{Lee1970}
B. H. Lee, J Appl. Phys. \textbf{41}, 2984 (1970).


\bibitem{Blattner1972}
see R. J. Blattner, L. K. Walford, and T. O. Baldwin,
J. Appl. Phys. 43, 935 (1972) for detailed references.



\bibitem{Landolt1999}
Landolt-B\"ornstein Tables, edited by  U. R\"osler, New Series Vol. III/41b (Springer, New York, 1999).


\end{thebibliography}
\end{document}